
\input jnl.tex
\input reforder.tex
\ignoreuncited
\singlespace
\title{QUANTUM CHAOS: A DECOHERENT DEFINITION.}
\author{Wojciech Hubert Zurek$^{(1)}$ and Juan Pablo Paz$^{(1,2)}$}
\affil{$(1)$: Theoretical Astrophysics, Los Alamos National Laboratory,
Los Alamos, NM 87545, USA}
\affil{$(2)$: Departamento de F\'\i sica, Facultad de Ciencias Exactas y
Naturales, Pabell\'on 1, Ciudad Universitaria, 1428 Buenos Aires,
Argentina}
\abstract{ We show that the rate of increase of von Neumann entropy
computed from the reduced density matrix of an open quantum system is
an excellent indicator of the dynamical behavior of its classical hamiltonian
counterpart. In decohering quantum analogs of systems which exhibit classical
hamiltonian chaos entropy production rate quickly tends to a constant which
is given by the sum of the positive Lyapunov exponents, and falls off only
as the system approaches equilibrium. By contrast, integrable systems
tend to have entropy production rate which decreases as $t^{-1}$ well before
equilibrium is attained. Thus, behavior of quantum systems in contact
with the environment can be used as a test to determine the nature of
their hamiltonian evolution.}

\head{1. Introduction}

Ever since the inception of quantum theory, the issue of the correspondence
between quantum and classical has been at the center of interest.
Perhaps the best known (and most durable!) problem
arises in the context of measurements\refto{WheelerZurek}.
The difficulty of representing the Universe as a whole -- including us,
the observers -- by means of quantum theory has been known to the forefathers
of quantum physics, and (in spite of the significant recent progress)
continues to be hotly debated\refto{Mazagon,Koln}. The problem of
the correspondence between the quantum and the classical in quantum
analogs of systems which classically exhibit dynamical chaos has come into
focus within the past two decades\refto{Chirikov,Hamiltonian,QuantumCh}.
It is perhaps best illustrated by the fact\refto{Casati} that, in course of
the hamiltonian evolution, quantum and classical versions of the same system
begin to exhibit significant discrepancies between the expectation values
of the
same quantity on a relatively short timescale:
$$
t_\chi\propto\lambda^{-1} \log(\chi \delta p/\hbar).\eqno(tchi)
$$
Above $\lambda$ is the Lyapunov exponent, $\chi$ is the scale
over which the potential becomes significantly nonlinear, and $\delta p$
characterizes the initial spread of the wavepacket in the units of momentum.

This timescale is only logarithmically dependent on the value of the
Planck constant $\hbar$. Thus, it is unconfortably short, even for
macroscopic systems: If one were to take this prediction on the face
value one would anticipate that in our Universe quantum chaotic systems
should stop obeying classical laws after a few dynamical timescales (which
is typically the order of the inverse Lyapunov exponent $\lambda^{-1}$).
By contrast, quantized regular (integrable) systems
tend to follow predictions of classical mechanics for much longer time
-- on a timescale of the order of $(1/\hbar)^\beta$ where $\beta$ is
some (positive) characteristic power. This failure of the correspendence
principle for chaotic systems has even led some to wonder whether
quantum theory can be the fundamental theory of our Universe which -- after
all -- seems to follow classical mechanics at the macroscopic
level\refto{Ford}.

The purpose of this paper is to provide a brief summary of the relevant
aspects of the process of decoherence -- which was
introduced\refto{Decoherence,JoosZeh} to deal with
the transition from quantum to classical in quantum measurements -- and
to show how it helps resolve the problem of quantum--classical correspondence
in the context of chaos. As an important corollary of this discussion we will
conjecture a simple method to diagnose chaos in the fully quantum system.
We shall base it on the behavior of the von Neumann entropy production rate of
systems coupled to an environment. In cases where the classical system
is chaotic, von Neumann entropy is conjectured to increase at a rate given
by Lyapunov exponents in its decohering quantum analog. By contrast, in quantum
analogs of regular systems entropy will grow at a rate which will
asymptotically
tend to zero well before the system reaches equilibrium.

What is remarkable about this result\refto{zp1} is that the entropy production
rate -- following the initial onset of decoherence which ocurrs on
a {\it decoherence timescale}\refto{Zurek86} which is essentially independent
of the system's self-hamiltonian, but dependent on the
strength and nature of the coupling with the environment\refto{phz} and
on the form of the initial state\refto{Decoherence,zhp} -- tends to be
dictated by the dynamics of the system rather than by the type or the strength
of the coupling. Independence from the strength and nature of the coupling
holds for a wide range of parameters in spite of the fact that the ultimate
cause of irreversibility is precisely the coupling with the environment.
This behavior mirrors classical intuition about
the nature of chaotic systems: Their evolution is -- in contrast
to regular systems -- unpredictable. We show that this unpredictability
carries over into the quantum domain, provided that systems which are
{\it open} are investigated. This result can be therefore regarded as
an additional indication that the correspondence between quantum and classical
dynamics should be sought only with the assistance of environment -- induced
superselection, the consequence of the process of
decoherence. Furthermore, entropy production rate in a decohering quantum
system can be regarded as a diagnostic tool: Rate of increase of entropy
can distinguish chaotic and regular quantum evolutions, thus providing a
completely quantum definition of quantum chaos.

\head{2. Decoherence}

Decoherence and its relation with quantum measurement are not the main subjects
of this paper. We shall summarize decoherence only very briefly with the eye
on its significance to the subject of the transition from quantum to classical
in the context of quantum chaos. A more complete review can be found
elsewhere\refto{zurekpt, zurek93}

Decoherence is the process of loss of (phase) coherence by the system caused
by the interaction with the external or internal degrees of freedom which
cannot be followed by the observer and are summarily called `the environment'.
Different states in the Hilbert space of the system of interest show
various degrees of susceptibility to decoherence. States which are least
susceptible (i.e., take longest to decohere) form the {\it preferred basis}
(also known as the {\it pointer basis}, in the context of quantum
measurement)\refto{Decoherence,zurekpt,zurek93,ZurekPT}.

Preferred states are singled out by the interaction between the system and
the environment. In idealized discussions of quantum apparatus,
complete immunity to decoherence can be guaranteed for the eigenstates of
the {\it pointer observable} which commutes with the total hamiltonian
(i.e., self hamiltonian plus the interaction hamiltonian). Hence, pointer
observable is conserved in spite of the interaction with the external degrees
of freedom\refto{Decoherence}.

Monitoring by the environment is a useful way of thinking about the emergence
of the preferred set of states and about the process of decoherence in
general. It can be shown that the interaction with the environment can be
regarded as a {\it continuous measurement} of the pointer observable. As
a consequence, the environment acquires a record of pointer observable. Its
states become correlated with the preferred pointer states: Quantum
state of the complete (system plus environment) object can be written as;
$$
|\Psi>=\sum_i\alpha_i |\sigma_i> |\epsilon_i> \ , \eqno(totalwf)
$$
where the states of the environment correlated to the eigenstates of the
preferred observable become (to an excellent approximation)
orthogonal (i.e., $<\epsilon_i|\epsilon_j>\propto\delta_{ij}$) as a result of
the interaction with the system.

As the environment continuously acquires the record of the states
$\{|\sigma_i>\}$, other states (linear superpositions of the preferred
states) are unstable. Quantum coherence of superpositions of $\{|\sigma_i>\}$
is lost: When a system is prepared in a state;
$$
|\Phi>=\sum_i\alpha_i |\sigma_i>, \eqno(systemwf)
$$
it will rapidly decay into a density matrix which is always diagonal in
the same (preferred) basis:
$$
|\Phi><\Phi| \ \rightarrow \
\rho\approx\sum_i|\alpha_i|^2|\sigma_i><\sigma_i|.\eqno(transit)
$$
Hence, the system behaves as if an effective superselection rule precluding
existence of superpositions between the eigenstates of the preferred
basis was in place.

{\it Environment induced superselection rules} effectively outlaw arbitrary
superpositions. Thus, even though the superposition principle is valid in
a closed quantum system, it is invalidated by decoherence for systems
interacting with their environments. All of the macroscopic quantum systems
we encounter in our everyday existence, as well as our own memory
and information processing hardware (e.g., neurons, etc) are macroscopic
enough and sufficiently strongly coupled to the environment to be
susceptible to decoherence, which will eliminate truly quantum superpositions
on a very short timescale. This process is absolutely essential in the
transition from quantum to classical in the context of quantum measurements
(where the classical apparatus tends to be very macroscopic) although
resolutions based on decoherence may not be easily palatable to everyone
(i.e., see comments on decoherence in the April 1993 issue of {\it Physics
Today}).

The timescale on which decoherence takes place can be estimated by solving
a specific example: a one dimensional particle moving in a potential
$V(x)$ coupled through its position with a thermal environment -- e.g.
with a collection of harmonic oscillators at a temperature $T$. Under
the appropriate assumptions one can derive an equation for the reduced density
matrix of the preferred particle. In the position representation
it reads \refto{QBM}:

$$\dot \rho \  = \
\overbrace{\underbrace{ -{i \over \hbar}
[H_o, \rho]}_{\dot p = - FORCE = \nabla V}}^{von~
Neumann~eq.} - \
\overbrace{\underbrace{\gamma (x-y) ({\partial \over \partial x}
\ - \  {\partial \over \partial y})\rho}_{\dot p=-\gamma p}}^{relaxation}
\  - \ \overbrace{\underbrace{{2 m \gamma k_B T \over \hbar^2} (x-y)^2 ~
\rho}}_{classical~phase~space}^{decoherence}
\eqno(mastereq)$$

Above, we
have indicated the role of the three terms which constitute the master equation
in the so--called
high temperature limit -- that is, in the case
when the thermal excitations of the environment dominate the effect of the
environment, and the effect of the vacuum fluctuations can be neglected.

Mathematically, classical limit is often associated with the size of Planck's
constant. In the limit $\hbar\rightarrow 0$ the very last term of the
above equation becomes dominant. To understand its effect, let us write
an explicit solution of \(mastereq) approximating the right hand side by the
last term only. In that case
$$
\rho(x,y,t)=\rho(x,y,0)\exp(-(x-y)^2D t/\hbar^2)\eqno(rhot)
$$
Above $D=2 m\gamma k_BT$.

It is now apparent that the evolution under nothing but the decoherence term
leaves the diagonal of the density matrix in the position representation
essentialy uneffected. By contrast, for $x\neq y$ the density matrix will decay
exponentially in a decoherence timescale\refto{Zurek86};
$$
\tau_D=\gamma^{-1} {\hbar^2\over D (x-y)^2}= \tau_R
({\lambda_{dB}\over\Delta x})^2 \ , \eqno(dectime)
$$
where $\lambda_{dB}=(\hbar^2/2mk_BT)^{1/2}$ is the thermal de Broglie
wavelength and $\tau_R=\gamma^{-1}$ is the relaxation timescale.

Two remarks are in order: (i) The decoherence timescale $\tau_D$ is much
shorter than the relaxation timescale $\tau_R$ for all macroscopic
situations, as typical thermal de Broglie wavelengths of macroscopic
bodies are many orders of magnitude smaller than macroscopic separations
$\Delta x$. (ii) The devastating effect of decoherence on superpositions
of position can be traced back to the preferential monitoring of that
observable ($x$) by the environment, which was coupled to the position of
the system of interest. This also tends to be the case in general: Interaction
potentials depend on position and, therefore, allow the environment to
monitor $x$\refto{Decoherence,phz,zurekpt,zurek93}. As a result of the action
of
the decoherence term, the vast majority of states which could in principle
describe the system of interest would be, in practice, eliminated by
the resulting environment - induced superselection. Only localized
states will be able to survive. They will form a preferred basis. For
(even though they will be in general still somewhat
unstable under the joint action of the self--hamiltonian and the
environment) they will be much more stable than their coherent superpositions.
This can be gauged by estimating the timescale characterizing the rate
of entropy production. For the preferred states this timescale will be
relatively long, determined by the dynamics and relaxation. For example,
in an underdamped harmonic oscillator the preferred states turn out to be
the familiar coherent states\refto{zhp}: Oscillator dynamics rotates all of
the states, which, in effect, translates spread in position into spread
in momentum (and vice versa) every quarter period of the oscillation.
As a result, coupling to position can be quite faithfully represented in the
``rotating wave approximation'' which makes the master equation symmetric in
$x$ and $p$ \refto{Louisell}.
Hence, coherent states will minimize entropy production. By contrast,
for superpositions of coherent states entropy production will happen on a very
much shorter decoherence timescale.

More general dynamics tends to deform states in the Hilbert space of the
system. Chaotic dynamics is especially effective in this, as it reflects
exponential stretching and squeezing of phase space distributions. Thus,
a regular patch in the phase space will be relatively quickly (on a
Lyapunov timescale) deformed into something which will be no longer
regular. By the same token, localized preferred states will tend to be
stretched into non--local superpositions. Thus, the form of a typical
state on the diagonal of the density matrix of a chaotic system will be
a matter of compromise between the chaotic dynamics and decoherence.
We shall discuss the nature of this compromise in the next section.

\head{3. Decoherence vs. exponential instability}

Phase space provides a natural arena to study the consequences of
the chaotic dynamics and its interplay with decoherence.
Master equation \(mastereq) can be translated into an equation for the
Wigner distribution. The resulting equation consists of the Wigner
transform of the commutator (which results in the so--called Moyal
bracket) plus additional terms representing relaxation and decoherence:
$$
\dot W=\{H,W\}_{MB}+2\gamma\partial_pp W + D\partial^2_{pp}W\eqno(wignereq)
$$
where the first term on the right hand side denotes the Moyal bracket, which
can be written in terms of the Poisson bracket as $\{\ ,\ \}_{MB}=2i\sin(\hbar
\{\ ,\ \}_{PB}/2i)/\hbar$.

We will be interested in the regime in which the coupling to the environment
is sufficiently weak so that the damping (represented by the second term
in \(wignereq)) is negligible. This is the so--called ``reversible classical
limit''\refto{Zurek86,zurekpt,phz} which in integrable systems yields
reversible classical trajectories but still eliminates non--local
superpositions (this limit is achieved by letting $\gamma$ approach zero
but keeping $D$ constant so that decoherence continues to be effective).
In this limit, and in the case where the potential is analytic, equation
\(wignereq) can be rewritten as\refto{Wigner}:
$$
\dot
W=\{H,W\}_{PB}+\sum_n{\hbar^{2n}(-1)^n\over(2n+1)!2^{2n}}\partial^{(2n+1)}_x V
\partial^{(2n+1)}_pW + D\partial^2_{pp}W.\eqno(wignereq2)
$$
Thus, Liouville flow in the phase space (and, therefore, classical dynamics) is
obtained from the basic quantum picture as long as the corrections appearing
in \(wignereq2) are negligible. However, in a chaotic system evolution of the
Wigner function generated by the Poisson bracket takes it quickly into the
regime where Poisson bracket alone does not suffice. This is because chaotic
systems exhibit exponential sensitivity to initial conditions. Consequently,
a phase space patch corresponding to a Wigner distribution will be
exponentially stretched in the unstable directions corresponding to
positive Lyapunov exponents. As the volume in phase space corresponding to
$W$ must be preserved, this will result in exponential shrinking
in other directions. Consequently, derivatives of the Wigner function with
respect to momentum (which enter
into the correction term) will exponentially increase, so that after a time
which is logarithmic in $\hbar$ these initially small terms will become
comparable to the Poisson bracket and Liouville dynamics will cease to be
an accurate approximation. This argument leads, in fact, to
a demonstration of equation \(tchi), as the reader is encouraged to verify.

One can regard this breakdown of the Liouville dynamics as a consequence of the
loss of validity of a classical formula for the force in terms of the gradient
of the potential $V(x)$ (which is implemented in the Poisson bracket). As
the Wigner function becomes more squeezed in momentum, by virtue of
Heisenberg's uncertainty principle it spreads in position, and it begins to
coherently sample increasingly large regions of the phase space.
This process results in the
domination of the evolution operator by the quantum forces when the extent
of the wavefunction in space becomes comparable with the scale of nonlinearity,
which for the various terms in equation \(wignereq2) is given by:
$$
\chi_n=(\partial_xV/\partial_x^{(2n+1)}V)^{1/2n}.\eqno(chin)
$$
How can decoherence help reestablish the quantum - classical correspondence?
Let us, for the moment, keep just the Poisson bracket and the diffusion term.
Then, in the neighbourhood of any point, equation \(wignereq2) can be easily
expanded along the unstable ($\lambda_i^+>0$) and stable ($\lambda_i^-<0$)
directions in phase space ($\sum_i(\lambda_i^-+\lambda_i^+)=0$).
Diffusion will have little influence on the evolution of $W$ along the
unstable directions: After the possible initial (decoherence timescale)
transient, $W$ will be stretched simply as a result of the dynamics, so
that the gradients along these directions will tend to decay anyway, without
assistance from diffusion. By contrast, squeezing which occurs along the
contracting directions will tend to be opposed by the diffusion. This will
lead to a steady state with the solution asymptotically approaching a Gaussian
with a half--width given by the critical dispersion:
$$
\sigma_{c_i}^2=2D_i/|\lambda_i^-|\eqno(sigmacrit)
$$
where $\lambda_i^-$ is the (negative) Lyapunov exponent along the stable
direction and $D_i$ is the diffusion coefficient along the same direction.
Below, we will assume that the diffusion is isotropic (as would be the
case in the rotating wave approximation). Thus, after some time (and in
the absence of folding -- the other aspect of chaos which we will discuss
below) the Wigner function will evolve into a multidimensional
``hyper--pancake,'' still stretching along the unstable directions but
with its width limited from below in the stable directions by equation
\(sigmacrit).

At this stage, entropy will be approximated by the logarithm of the
effective volume of the hyper--pancake. As its extent
in the stable direction is fixed by the critical width \(sigmacrit), its
volume will tend to increase at a rate given by the positive exponents.
Consequently,
$$\dot H \approx \sum_i\lambda_i^+.\eqno(hdot)
$$
This constant rate will set in after a time larger than the decoherence
timescale (for smaller times the entropy production can be even more rapid)
and after a time over which the initial Wigner distribution becomes squeezeed
by the dynamics to the dimension of order of the critical dispersion
$\sigma_{c_i}$. Equation \(hdot) will be valid untill the pancake fills in
the available phase space and the system reaches (approximately) uniform
distribution ove the accessible part of the phase space, that is after
a time defined by;
$$
t_{eq}=(H_{eq}/H_0)/\dot H,\eqno(eqtime)
$$
where $H_0$ is the initial entropy, and $H_{eq}$ is the entropy uniformized
by the chaotic dynamics.

Astute reader will note that $H_{eq}$ above need not be a true equilibrium
entropy with the temperature given by $T$. Rather, it will correspond to
dynamical quasi--equillibrium --
the approximately uniform distribution over this part of the phase space
which is accessible to the chaotic system as a result of its dynamics.

The corresponding timescale will have a similar dependence on
$\hbar$ as the timescale $t_{\chi}$ defined by Eq. (1). This is because entropy
is approximately given by the logarithm of the volume of the phase space over
which the probability distribution has spread in the units of Planck constant.
Nevertheless, $t_{\chi}$ and $t_{eq}$ depend on rather different aspects of the
initial and final state, and one can expect $t_{\chi}$ to be be typically
a fraction of $t_{eq}$.

By contrast, in integrable systems stretching of
the corresponding hyper--pancake in
phase space will proceed only polynomially. Thus, even when it will get to
the stage at which, in the contracting direction, diffusion will become
important, stretching in the unstable direction will be only polynomial
(rather than exponential). Consequently, the volume of the hyper--pancake
will increase only as some power of time. Hence, the entropy will grow
only logarithmically as the entropy production rate will fall as
$\dot H\propto 1/t$: It will take exponentially long to approach dynamical
quasi--equilibrium.

This difference in behavior between chaotic and integrable {\it open}
quantum systems is striking and can be used as a defining feature of
quantum chaos.

\head{4. Quantum--classical correspondence in chaotic systems. }

The failure of Ehrenfest theorem in chaotic systems is the consequence of
the exponentially unstable Liouville flow which compresses Wigner function
into an exponentially narrowing pancake. As the momentum becomes
progressively squeezed -- which makes it less and less uncertain -- the
spatial extent of the coherent quantum wavefunction will exponentially
increase until it eventually simultaneously samples much of the
potential well. By then the force is no longer given by a gradient of the
potential: The wavefunction is too non--local for such a formula. It would
not be even clear {\it where} (whithin the spatial support of the
wavefunction) one should compute such gradient. Instead, a more complicated
formula, the Moyal bracket, is needed.

Decoherence limits the extent over which the wavefunction can remain
coherent. This is because a finite minimal dispersion in momentum
\(sigmacrit) corresponds to quantum coherence over distances no longer than:
$$
l=\hbar/\sigma_c=\hbar(2D/\lambda)^{-1/2}.\eqno(lcrit)
$$
Thus, when the scale \(chin) on which nonlinearities in the potential are
significant is small compared to the extent of the wavefunction
$$
\chi\ll l\eqno(chivsl)
$$
decoherence will have essentially no effect. Evolution will remain purely
quantum and will be generated by the full Moyal bracket.

By contrast, when the opposite is true, the evolution will never squeeze
Wigner distribution function enough for the full Moyal bracket to be
relevant. Poisson bracket will suffice to approximate the flow of probability
in phase space. The inequality characterizing this case can be written in
a manner reminiscent of the Heisenberg indeterminacy principle:
$$
\hbar\ll \chi\sigma_c.\eqno(ineqclass)
$$
That is, as long as decoherence keeps the state vector from becoming too
narrow in momentum, it will also prevent it from sampling the potential
coherently over distances on which $V(x)$ is noticeably different from linear.
Hence, local gradients will suffice in the evaluation of the forces --
Poisson bracket is all that is required.

There is one more interesting regime where the chaotic motion is
dynamically reversible (that is, $\dot H=0$) even if the system satisfies
inequality \(ineqclass). This happens when the initial patch in phase
space is large (volume much larger than the Planck volume -- initial
entropy larger than a single bit) and regular. Then the initial stage
of the evolution will proceed reversibly, in accord with the Poisson
bracket generated flow. Decoherence will have little effect. This is
because its influence will set in only as the dimension of the Wigner
distribution in the contracting dimension will approach the critical
dispersion $\sigma_c$: In a simple example the entropy production will
increase as:
$$
\dot H = \lambda{1\over \Bigl(1+({\sigma_p^2(0)\over\sigma_c^2}-1)\exp(
-2\lambda t)\Bigr)}\eqno(hdotapp)$$

So far, we have not taken into account (or, at least, not taken into
account explicitelly) the other major characteristic of chaos: In addition
to exponential instability, chaotic systems ``fold'' the phase space
distribution. While this problem may require further study, we believe that
the fundamentals of folding are already implicit in the above discussion:
Folding will happen on the scale $\chi$ of nonlinearities in the
potential (which will typically -- but not always -- coincide with the size of
the system, as it is defined by the range of its classical trajectory).
Hence, preventing the system from maintaining coherence over distances
of the order of $\chi$ will also ascertain its classical behavior in course
of folding. There will simply be no coherence left between the fragments
of the wavepacket which will come into proximity as a result of folding, if
they had to be separated by distances larger than $l$ in the course of the
preceding evolution. Thus, folding will proceed as if the system was
classical, but with a proviso: After sufficiently many folds the distribution
function (which in the stable direction cannot shrink to less than $\sigma_c$)
will simply fill in the available phase space. This will be achieved in
the previously defined equilibrium timescale $t_{eq}$. These conclusions are
consistent with the numerical studies of quantum maps corresponding to open
quantum systems such as the ``standard map'' carried out by Graham and
his coworkers\refto{Graham}.

\head{5. Summary.}

We have used the entropy production rate in a decohering quantum system
to characterize the nature of its evolution. Classical unpredictability
-- the essence of dynamical chaos -- was shown to beget quantum
unpredictability, quantified by the rapid entropy production on the
Lyapunov timescale. By contrast, dynamics of integrable systems leads to
only gradual (polynomial) spread of the patch of the phase space corresponding
to the state of the system. As a result, a much slower evolution towards
dynamical quasi--equilibrium (and a relatively good predictability in spite
of the coupling to the environmnet) characterize quantum analogs of classically
integrable systems. This distinction is conjectured to be a diagnostic of
the dynamical nature of the system.

\references

\refis{WheelerZurek} J. A. Wheeler and W. H. Zurek, {\it Quantum Theory
and Measurement}, (Princeton University Press, Princeton, 1983).

\refis{Mazagon} J. J. Halliwell, J. Perez-Mercader, and W. H. Zurek, eds., {\it
Physical Origins of Time Asymmetry}, (Cambridge University Press, 1994).

\refis{Koln} P. Busch, P. Lahti, and P. Mittelstaedt, eds., {\it Quantum
Measurements,
Irreversibility, and the Physics of Information} (World Scientific, Singapore,
1994).

\refis{zurekpt} W. H. Zurek, Physics Today {\bf 44}, 36 (1991).

\refis{zurek93} W. H. Zurek, Prog. Theor. Phys. {\bf89}, 281 (1993).

\refis{zhp} W. H. Zurek, S. Habib and J.P. Paz, Phys. Rev. Lett. {\bf 70}, 1187
(1993).

\refis{zp1} W. H. Zurek and J. P. Paz, Phys. Rev. Lett. {\bf 72}, 2508 (1994).

\refis{phz} J. P. Paz, S. Habib and W. H. Zurek,  Phys. Rev. {\bf D 47} (1993)
488.

\refis{qbm} B. L. Hu, J. P. Paz and Y. Zhang, Phys. Rev. {\bf D 45},
(1992) 2843; {\it ibid} {\bf D 47} (1993) 1576.

\refis{Chirikov} B.V. Chirikov, Phys. Rep. {\bf 52}, 263 (1979); G. M.
Zaslavsky, Phys. Rep.
{\bf 80}, 157 (1981); M--J. Giannoni, A. Voros, J. Zinn--Justin, eds., {\it
Chaos and Quantum
Physics}, Les Houches Lectures LII (North Holland, Amsterdam, 1991).

\refis{Ford} J. Ford and G. Mantica, Am. J. Phys. {\bf 60}, 1086 (1992).

\refis{Hamiltonian} See, e.g. G. Ioos, R. H. G. Hellmann and R. Stora eds.,
{\it Chaotic Behavior in
Deterministic Systems}, Les Houches Lectures XXXVI (North Holland, Amsterdam,
1983);
R. S. Mc Kay and J. O. Meiss eds., {\it Hamiltonian Dynamical Systems} (Hilger,
Philadelphia, 1987).

\refis{QuantumCh} M. B. Berry, Proc. Roy. Soc. London, {\bf A413}, 183
(1987); M. Gutzwiller,
{\it Chaos in Classical and Quantum Mechanics} (Springer Verlag, New York,
1990);
F. Haake, {\it Quantum Signature of Chaos} (Springer Verlag, New York, 1990)
and references therein.

\refis{Decoherence} W. H. Zurek, Phys. Rev. {\bf D24}, 1516 (1981); {\it ibid},
{\bf D26}, 1862 (1982).

\refis{JoosZeh} E. Joos and H. D. Zeh, Zeits. Phys. {\bf B59}, 229 (1985).

\refis{ZurekPT} W. H. Zurek, Physics Today, {\bf 46}, 81 (1993).

\refis{PazHZ} J. P. Paz, S. Habib and W. H. Zurek, Phys. Rev. {\bf D47}, 488
(1993).

\refis{ZurekHP} W. H. Zurek, S. Habib and J. P. Paz, Phys. Rev. Lett. {70},
1187 (1993).

\refis{Zurek86} W. H. Zurek, pp. 145-149 in {\it Frontiers of Nonequilibrium
Statistical Mechanics}, G.
T. Moore and M. O. Scully eds., (Plenum, New York, 1986).

\refis{GMH} M. Gell--Mann and J. B. Hartle, Phys. Rev. {\bf D47}, 3345 (1993).

\refis{QBM} A. O. Caldeira and A. J. Leggett, Physica {\bf 121A}, 587 (1983);
W.
G. Unruh and W. H. Zurek, Phys Rev. {\bf D40}, 1071 (1989); B. L. Hu, J. P. Paz
and Y. Zhang, Phys. Rev. {\bf D45}, 2843 (1992); {\it ibid} {\bf D47}, 1576
(1993).

\refis{Louisell} W. H. Louisell, {\it Quantum Statistical Properties of
Radiation} (Wiley, New
York, 1973);
V. Buzek and P. Knight; {\it Quantum interference, superposition states of
light
and nonclassical effects}, Prog. in Opt. (1993).

\refis{vonNeumann} J. von Neumann, {\it Mathematical Foundations of Quantum
Mechanics},
english translation by R. T. Beyer (Princeton Univ. Press, Princeton, 1955); H.
D. Zeh,
{\it The Direction of Time}, (Springer--Verlag, New York, 1991).

\refis{Blatt} J. M. Blatt expressed a similar suspicion in a compelling,
although less rigorous
and completely classical paper, Progr. Theor. Phys. {\bf 22}, 745 (1959)

\refis{Milburn} G.J. Milburn and C.A. Holmes, Phys. Rev. Lett. {\bf 56}, 2237
(1986).

\refis{Casati} G. P. Berman and G. M. Zaslawsky, Physica {\bf A91}, 450 (1978);
G. Casati, B. V. Chirikov, F. M. Izraileev and J. Ford, {\it Lectures Notes in
Physics} {\bf 93}, (Springer--Verlag, New York, 1979).

\refis{Graham} T. Dittrich and R. Graham, Z. f\"ur Phys. {\bf B62},
515 (1986); T. Dittrich and R. Graham, Ann. Phys. (NY) {\bf 200}, 363 (1990);
T. Dittrich and R. Graham, ``Quantum chaos in open systems'', in
{\it Information Dynamics}, H. Atmanspacher and H. Scheingraber, eds.,
NATO ASI Series, {\bf B256}, 289, (Plenum, 1990) and
references therein.

\refis{Wigner} M. Hillery, R.F. O'Connell, M.O. Scully and E. Wigner, Phys.
Rep. {\bf 106}, 121 (1984).

\endreferences
\vfill\eject

Figure 1: The evolution of a ``patch'' in the phase space in a system with
an exponential instability:

(a) The case when the system is isolated and
only exponential stretching in the unstable direction as well as
the corresponding shrinking in the complementary direction take place.
The decrease of the dimension of the patch in momentum results
(through the Heisenberg indeterminacy relation) in nonlocality, which
leads to non--classical corrections to the expression for the force
(resulting in the Moyal bracket).

(b) When the system is open, decoherence prevents the dispersion of momentum
from shrinking to less than te critical dispersion $\sigma_c$. Critical
dispersion characterizes the steady state set by the competition between
the dynamics (which attempts to narrow the patch, as it was shown in Fig. 1a)
and decoherence, which is associated with the diffusion operator (which
attempts to spread the patch). When decoherence is sufficiently effective,
the spatial extent of the coherence of the wavepacket (given by
$l = \hbar/\sigma_c$, \(lcrit)) will be sufficiently small so that the
harmonic approximation to the potential will be accurate. And in such linear
regime Moyal bracket and Poisson bracket coincide. Therefore, classical
dynamics can be recovered even for chaotic systems.

\vfill\eject
\end